# Factorial survival analysis for treatment effects under dependent censoring

Takeshi Emura[1]  (corresponding author), Marc Ditzhaus[2], Dennis Dobler[3], Kenta Murotani[4]

**Abstract**

Factorial analyses offer a powerful nonparametric means to detect main or interaction effects among multiple treatments. For survival outcomes, e.g. from clinical trials, such techniques can be adopted for comparing reasonable quantifications of treatment effects. The key difficulty to solve in survival analysis concerns proper handling of censorings. So far, all existing factorial analyses for survival data were developed under the independent censoring assumption, which is too strong for many applications. As a solution, the central aim of this article is to develop new methods in factorial survival analyses under quite general dependent censoring regimes. This will be accomplished by combining existing results for factorial survival analyses with techniques developed for survival copula models. As a result, we will present an appealing *F*-test that exhibits sound performance in our simulation study. The new methods are illustrated in a real data analysis. We implement the proposed method in an R function *surv.factorial*(.) in the R package *compound.Cox*.

## 1. Introduction

Factorial designs allow researchers to assess treatment effects in one-way, two-way, and other layouts of experimental factors[1]. Factorial analyses offer a variety of treatment contrasts involving multiple treatment factors and their levels. The classical analysis of variance (ANOVA) is a test for detecting treatment effects via the *F*-statistic under normally distributed outcomes. The *F*-test is an *omnibus* test[2,3] for the global null hypothesis, where any significance indicates the presence of treatment effects, providing scientific evidence for further explorations of individual factors/treatments.

For survival outcomes, there is a keen interest for adopting factorial analyses for testing treatment effects in cancer clinical trials: see for instance, the ongoing phase III clinical trial with a $2 \times 2$ factorial design[4]. The 2017 survey identified 30 clinical trials with the $2 \times 2$ factorial design for phase III cancer treatment[5], including a trial that examined the effect of treatments on survival for patients with advanced cervical cancer[6].

---

[1] Biostatistics Center, Kurume University, 67 Asahi-machi, Kurume, JAPAN
   Research Center for Medical and Health Data Science, The Institute of Statistical Mathematics, JAPAN
[2] Faculty of Mathematics, Otto-von-Guericke University, Magdeburg, GERMANY
[3] Department of Mathematics, Faculty of Science, Vrije Universiteit Amsterdam, Amsterdam, NETHERLANDS
[4] Biostatistics Center, Kurume University, 67 Asahi-machi, Kurume, JAPAN

Due to the non-normality of survival data, treatment effects are formulated nonparametrically via the Mann-Whitney effects[7], pairwise effects using sample partitions[8], the effects on median survival[9], or the effects on cumulative hazards[10]. While these metrics provide different survival aspects of treatment effects, they can be identified without aid of specific distributions. These techniques offer a powerful nonparametric means to detect main/interaction effects of treatments.

The key difficulty to solve in survival analysis concerns proper handling of censorings owing to the limited follow-up time of clinical trials[11]. However, the existing factorial analyses[7-10]. were developed under the assumption that censoring mechanism is independent of survival. Unfortunately, this assumption is imposed for the sake of mathematical convenience, and its validity in real applications is largely questionable[12-18].

Dependent censoring arises in analysis of survival data, especially in data obtained from biomedical studies [14,19-29]. The Kaplan-Meier (KM) estimator for survival function is systematically biased when censoring is associated with survival time[30]. Andersen and Perme[19] suggested reducing the bias of the KM estimator with aid of observed covariates. Emura and Chen[16] analyzed the bias of Cox regression and variable selection under dependent censoring. Emura and Hsu[31] studied the bias on two-sample tests based on the Mann-Whitney type treatment effects. Other statistical tools for dependent censoring are reviewed by the books of Collett[20] and Emura and Chen[17]. However, dependent censoring has not been studied in the context of factorial designs.

In this article, we propose a new method for factorial survival analyses, which can take into account dependent censoring. Our method extends the factorial analysis of Dobler and Pauly[7] to the copula-based factorial analysis under dependent censoring. We employ copula models to account for dependent censoring as in Emura and Hsu[31] who worked on a two-sample test. The proposed method leads to an omnibus test for the global null hypothesis and other tests for composite null hypotheses under factorial designs under dependent censoring. The method also allows researchers to conduct sensitivity analyses for treatment effects under a variety of copula models for dependent censoring. We implement the method in an R function surv.factorial(.) in the R package *compound.Cox*[32].

This article is organized as follows. Section 2 reviews the background for factorial survival analyses. Section 3 proposes a new method for factorial survival analyses under dependent censoring. Section 4 explains the software implementation of the proposed method. Section 5 conducts simulation studies to check the performance of the proposed method. Section 6 provides a real data example. Section 7 concludes the article.

## 2. Background

We introduce the survival analysis method of factorial designs, which was proposed by Dobler and Pauly[7]. This method defines treatment effects and test them by using censored survival data collected from one-way, two-way, or other layouts.

### 2.1 Treatment effects

In clinical trials and animal experiments, the effects of treatments are often assessed by survival outcomes. The classical ANOVA for testing the equality of normal means is inappropriate since survival data are non-normally distributed. Therefore, the nonparametric formulations of treatment effects are needed[7,33].

Consider patients allocated to $d$ different treatments. Let $n_i$ be the sample size in the $i$-th treatment for $i = 1, 2, \ldots, d$. Define the survival function $S_i(t) = P(T_{i1} > t)$ for survival time $T_{ij}$ for $j = 1, 2, \ldots, n_i$. Let $N = \sum_{i=1}^{d} n_i$ be the total sample size. The *pairwise effect* comparing the $i$-th group with the $\ell$-th group is

$$w_{i\ell} = P(T_{i1} > T_{\ell 1}) + \frac{1}{2} P(T_{i1} = T_{\ell 1}) = -\int S_i^{\pm}(t) dS_\ell(t), \qquad (1)$$

where $S_i^{\pm}(t) = \{S_i(t+) + S_i(t-)\}/2$, $S_i(t+) = \lim_{s \downarrow t} S_i(s)$, and $S_i(t-) = \lim_{s \uparrow t} S_i(s)$. This is the probability that a patient in the $i$-th group survives longer than a patient in the $\ell$-th group. The case of $w_{i\ell} > 1/2$ (or $w_{i\ell} < 1/2$) gives a beneficial (or harmful) treatment for the $i$-th group relative to the $\ell$-th group. For $d = 2$, this pairwise effects of treatments have been useful in two-sample comparisons[34-38]. However, in factorial analyses involving multiple groups, the pairwise effects have to be aggregated to characterize the relative treatment effects.

Following the framework of the nonparametric ANOVA[33,39], Dobler and Pauly[7] defined the *relative treatment effect* on the $i$-th group relative to the *average treatment effect* as

$$p_i = \frac{1}{d} \sum_{\ell=1}^{d} w_{i\ell} = -\int S_i^{\pm}(t) d\bar{S}(t),$$

where $\bar{S} = \sum_{\ell=1}^{d} S_\ell / d$. One can write

$$\boldsymbol{p} = \begin{pmatrix} p_1 \\ \vdots \\ p_d \end{pmatrix} = \boldsymbol{A}\boldsymbol{w}, \qquad \boldsymbol{w} = (w_{11}, \ldots, w_{1d} \vdots w_{21}, \ldots, w_{2d} \vdots \cdots \vdots w_{d1}, \ldots, w_{dd})' \qquad (2)$$

where $A = I_d \otimes (1'_d/d)$, $I_d$ is the $d \times d$ identity matrix, $1'_d = (1,\ldots,1)$ is a $d$-vector of ones, and $\otimes$ is the Kronecker product. Hence, the problem of estimating the treatment effects $p$ reduces to the problem of estimating the pairwise effects $w$.

A variety of hypotheses can be formulated using a contrast matrix $C$ such that

$$H_0: Cp = 0 \quad \text{vs.} \quad H_1: Cp \neq 0.$$

The simplest example is $C = P_d \equiv I_d - 1_d 1'_d/d$, yielding the global hypothesis

$$H_0: p_1 = p_2 = \cdots = p_d \quad \text{vs.} \quad H_1: p_i \neq p_\ell, \exists (i, \ell).$$

Local hypotheses (e.g., $H_0: p_1 = p_d$) can also be formulated by appropriately specifying $C$.

For a two-way layout, we assume "Factor A" with $a$ levels and "Factor B" with $b$ levels. The treatment effects with interactions are

$$p = (p_1, p_2, \ldots, p_{ab})' = (p_{11}, \ldots, p_{1b} \vdots p_{21}, \ldots, p_{2b} \vdots \cdots \vdots p_{a1}, \ldots, p_{ab})',$$

where $p_{jk}$ is the treatment effect for the $j$-th level in Factor A and the $k$-th level in Factor B. For instance, the test for main effects in Factor A is

$$H_0: p_{1\cdot} = p_{2\cdot} = \cdots = p_{a\cdot} \quad \text{vs.} \quad H_1: p_{i\cdot} \neq p_{\ell\cdot} \text{ for } \exists (i, \ell), i \neq \ell,$$

where $p_{i\cdot} = \sum_{\ell=1}^{b} p_{i\ell}$. This hypothesis is specified by $C = P_a \otimes 1'_b/b$. Other hypotheses can be formulated by choosing appropriate contrast matrices, such as $1'_a/a \otimes P_b$ for main effects in Factor B and $P_a \otimes P_b$ for interactions.

## 2.2 Factorial analysis under independent censoring

We review the test for $H_0: Cp = 0$ under independent censoring, which was proposed by Dobler and Pauly[7]. The test rejects $H_0$ if $C\hat{p}$ deviates from $0$, where $\hat{p} = A\hat{w}$, and $\hat{w}$ is a $d^2$-vector whose element is

$$\hat{w}_{i\ell} = -\int \hat{S}_i^{\pm}(t) d\hat{S}_\ell(t),$$

where $\hat{S}_i(.)$ is the KM estimator for $S_i(.)$, and $\hat{S}_i^{\pm}(t) = \{\hat{S}_i(t+) + \hat{S}_i(t-)\}/2$.

Let $T = C'(CC')^+ C$ be a projection matrix, where $(CC')^+$ is the Moore-Penrose inverse of $CC'$ that is a unique characterization of the generalized inverse. The $F$-statistic is defined as

$$F_N = \frac{N\hat{p}'T\hat{p}}{\text{tr}(T\hat{V})} = \frac{(C\hat{p})'(CC')^+(C\hat{p})}{\text{tr}[(CC')^+\hat{V}CC']},$$

where $\hat{V}$ is the estimator of the asymptotic covariance of $\sqrt{N}(\hat{p} - p)$.

The null distribution of $F_N$ was derived by Dobler and Pauly[7]; $F_N$ converges to a distribution having the unit mean under $H_0: Cp = 0$. Under the alternative hypothesis $H_1: Cp \neq 0$, $F_N$ goes to infinity in probability. Therefore, a consistent test can be constructed by rejecting $H_0: Cp = 0$ for $F_N > c_{N,\alpha}$, where $c_{N,\alpha}$ is the upper $\alpha \times 100$ percent point of the null distribution. Dobler and Pauly[7] suggested calculating the value $c_{N,\alpha}$ via a multiplier bootstrap method. However, the test is valid only under the independent censoring assumption as the KM estimator is inconsistent under dependent censoring.

In this context, our goal is to modify $F_N$ to produce a consistent test under dependent censoring.

## 3. Proposed method under dependent censoring

We first formulate a survival copula model for dependent censoring. Then, we propose a new estimator of treatment effects and a new *F*-test for treatment effects under dependent censoring.

### 3.1 Preliminary: estimating survival under dependent censoring

Recall that survival time $T_{ij}$ may be right-censored by censoring time $U_{ij}$. Observed data consist of $\{(X_{ij}, \delta_{ij}); i = 1, \dots, d, j = 1, \dots, n_i\}$, where $X_{ij} = \min(T_{ij}, U_{ij})$ and $\delta_{ij} = 1\{T_{ij} \leq U_{ij}\}$, where $1\{.\}$ is the indicator function.

In order to model the structure of dependent censoring, we postulate the survival copula model:

$$P(T_{ij} > t, U_{ij} > u) = C_{i\theta_i}\{S_i(t), G_i(u)\}, \quad i = 1, 2, \dots, d, \tag{3}$$

where $C_{i\theta_i}(\cdot,\cdot)$ is a copula[40,41] with the parameter $\theta_i$, and $S_i(t) = P(T_{ij} > t)$ and $G_i(u) = P(U_{ij} > u)$ are survival functions. If the Clayton copula[42] is specified,

$$C_{i\theta_i}(u, v) = (u^{-\theta_i} + v^{-\theta_i} - 1)^{-\frac{1}{\theta_i}}, \quad \theta_i > 0, \quad i = 1, 2, \dots, d,$$

where the parameter $\theta_i$ is related to Kendall's tau for $T_{ij}$ and $U_{ij}$ through $\tau_{\theta_i} = \theta_i/(\theta_i + 2)$. If $C_{i\theta_i}(u, v) = uv$ were imposed for $i = 1, 2, \dots, d$, the model would reduce to the independent censoring model. Other examples of copulas can be seen in Appendix A.1. In many copulas, the limit $\theta_i \to 0$ leads to $C_{i\theta_i}(u, v) \to uv$. Thus, the model (3) provides a quite general dependent censoring regimes, including the independent censoring model as its special case of $\theta_i = 0$.

One can estimate $S_i(t)$ using observed data under the model (3). Specifically, we introduce the copula-graphic (CG) estimator[43], in particular, a version of the CG estimator derived by Rivest and Wells[30] for the subclass of the model (1) given by Archimedean copulas:

$$P(T_{ij} > t, U_{ij} > u) = \phi_{i\theta_i}^{-1}[\phi_{i\theta_i}\{S_i(t)\} + \phi_{i\theta_i}\{G_i(u)\}], \quad i = 1, \dots, d, \qquad (4)$$

where $\phi_{i\theta_i}$ is a generator function that is continuous and strictly decreasing from $\phi_{i\theta_i}(0) = \infty$ to $\phi_{i\theta_i}(1) = 0$[40]. Then, assuming that the true $S_i(t)$ and $G_i(u)$ are continuous (and hence, there are no ties in $X_{ij}$s), the CG estimator of Rivest and Wells[30] is derived as

$$\hat{S}_i^{CG}(t) = \phi_{i\theta_i}^{-1}\left[\sum_{j:\, X_{ij} \le t,\, \delta_{ij}=1} \left\{\phi_{i\theta_i}\left(\frac{\bar{Y}_i(X_{ij}) - 1}{n_i}\right) - \phi_{i\theta_i}\left(\frac{\bar{Y}_i(X_{ij})}{n_i}\right)\right\}\right], \quad 0 \le t \le \max_j(X_{ij}),$$

where $\bar{Y}_i(u) = \sum_{j=1}^{n_i} 1\{X_{ij} \ge u\}$ is the number at-risk at $u$. The estimator $\hat{S}_i^{CG}(.)$ is uniformly consistent for $S_i(.)$ when the model (4) is correct: see Theorem 1 of Rivest and Wells[30].

Under the Clayton copula, the generator function is $\phi_{i\theta_i}(t) = (t^{-\theta_i} - 1)/\theta_i$ for $\theta_i > 0$. Then, the CG estimator is

$$\hat{S}_i^{CG}(t) = \left[1 - \sum_{j:\, X_{ij} \le t,\, \delta_{ij}=1} \left\{\left(\frac{\bar{Y}_i(X_{ij}) - 1}{n_i}\right)^{-\theta_i} - \left(\frac{\bar{Y}_i(X_{ij})}{n_i}\right)^{-\theta_i}\right\}\right]^{-\frac{1}{\theta_i}}, \quad 0 \le t \le \max_j(X_{ij}).$$

Here, the parameter $\theta_i$ must be prespecified since it is difficult to estimate from data[44]. A common strategy is a sensitivity analysis that tries different values of $\theta_i$ and compares the results[30,31,45,46].

Note that the CG estimator reduces to the KM estimator under the independence copula given by $\phi_{i\theta_i}(t) = -\log(t)$. In this case, one has

$$\hat{S}_i^{CG}(t) = \hat{S}_i(t) = \prod_{j:\, X_{ij} \le t,\, \delta_{ij}=1}\left[1 - \frac{1}{\bar{Y}_i(X_{ij})}\right], \quad 0 \le t \le \max_j(X_{ij}).$$

The CG estimators under the Gumbel and Frank copulas are given in Appendix A.2.

**3.2 Proposed estimator for treatment effects**

We shall derive a new estimator for the relative treatment effects $\boldsymbol{p}$.

We first propose to estimate the pairwise effect $w_{i\ell}$ in Eq. (1) by

$$\widehat{w}_{i\ell}^{CG} = -\int \hat{S}_i^{CG\pm}(t) d\hat{S}_\ell^{CG}(t).$$

Then, by $\boldsymbol{p} = \boldsymbol{A}\boldsymbol{w}$ of Eq. (2), we propose our estimator of $\boldsymbol{p}$ as

$$\widehat{\boldsymbol{p}}^{CG} = \begin{pmatrix} \hat{p}_1^{CG} \\ \vdots \\ \hat{p}_d^{CG} \end{pmatrix} = \boldsymbol{A}\widehat{\boldsymbol{w}}^{CG} = \boldsymbol{A}(\widehat{w}_{11}^{CG}, \ldots, \widehat{w}_{1d}^{CG} \vdots \widehat{w}_{21}^{CG}, \ldots, \widehat{w}_{2d}^{CG} \vdots \cdots \vdots \widehat{w}_{d1}^{CG}, \ldots, \widehat{w}_{dd}^{CG})'.$$

Under certain conditions, the estimator $\widehat{\boldsymbol{p}}^{CG}$ is consistent for $\boldsymbol{p}$, and permits the asymptotic normal approximation $\sqrt{N}(\widehat{\boldsymbol{p}}^{CG} - \boldsymbol{p}) \sim N_d(\boldsymbol{0}, \boldsymbol{V})$, $\boldsymbol{V} = \boldsymbol{A}\boldsymbol{\Omega}\boldsymbol{A}'$, where $\boldsymbol{\Omega}$ is defined in Appendix A.3. The proof of the asymptotic normality is given in Appendix A.3.

The variance of $\widehat{\boldsymbol{p}}^{CG}$ needs to be estimated to construct a test statistic. However, as the forms of $\boldsymbol{\Omega}$ in the asymptotic variance is complex, we suggest applying a jackknife estimator of variance[47]. Following Emura and Hsu[31], we estimate the variance of $\widehat{\boldsymbol{p}}^{CG}$ by

$$\boldsymbol{V}(\widehat{\boldsymbol{p}}^{CG}) = \begin{pmatrix} \hat{\sigma}_1^2 & \cdots & \hat{\sigma}_{1d} \\ \vdots & \ddots & \vdots \\ \hat{\sigma}_{d1} & \cdots & \hat{\sigma}_d^2 \end{pmatrix} = \frac{N}{N-1} \sum_{i=1}^{d} \sum_{j=1}^{n_i} (\widehat{\boldsymbol{p}}^{CG(-ij)} - \widehat{\boldsymbol{p}}^{CG(.)})(\widehat{\boldsymbol{p}}^{CG(-i\ )} - \widehat{\boldsymbol{p}}^{CG(.)})',$$

where $\widehat{\boldsymbol{p}}^{CG(-ij)}$ is computed without the $j$-th patient in the $i$-th treatment, and $\widehat{\boldsymbol{p}}^{CG(.)} = \sum_{i=1}^{d} \sum_{j=1}^{n_i} \widehat{\boldsymbol{p}}^{CG(-ij)}/N$. The standard error (SE) is $\text{SE}(\hat{p}_i^{CG}) = \hat{\sigma}_i$, and the $(1-\alpha) \times 100 \%$ confidence interval (CI) for $p_i$ is

$$[\hat{p}_i^{CG} - z_{\alpha/2} \text{ SE}(\hat{p}_i^{CG}), \quad \hat{p}_i^{CG} + z_{\alpha/2} \text{ SE}(\hat{p}_i^{CG})].$$

If necessary, the CI is transformed into [0, 1] by the logit function or others[7,33,48]. Note that the bootstrap resampling method does not work since the CG estimator does not adopt to tied data.

## 3.3 The proposed test

We shall propose a new test for a linear hypothesis $H_0: \boldsymbol{Cp} = \boldsymbol{0}$ vs. $H_1: \boldsymbol{Cp} \neq \boldsymbol{0}$ under factorial designs, where $\boldsymbol{C}$ is a contrast matrix. The test detects the departure of $\boldsymbol{C\hat{p}}^{CG}$ from $\boldsymbol{0}$. In the following, we shall extend the so-called "ANOVA-type statistic"[7,33,39] for survival analysis with dependent censoring.

Let $\boldsymbol{T} = \boldsymbol{C}'(\boldsymbol{CC}')^+\boldsymbol{C}$ be a projection matrix. The $F$-statistic is defined as

$$F_N^{CG} = \frac{N\hat{\boldsymbol{p}}^{CG\prime}\boldsymbol{T}\hat{\boldsymbol{p}}^{CG}}{\mathrm{tr}(\boldsymbol{T}\hat{\boldsymbol{V}}^{CG})},$$

where $\hat{\boldsymbol{V}}^{CG} = N \times \boldsymbol{V}(\hat{\boldsymbol{p}}^{CG})$ is the jackknife estimator of the asymptotic covariance of $\sqrt{N}(\hat{\boldsymbol{p}}^{CG} - \boldsymbol{p})$. Then, under $H_0$, the distribution of $F_N^{CG}$ is approximated by a weighted chi-squared distribution,

$$F_N^{CG} \sim \frac{\sum_{i=1}^{d}\lambda_i \chi_i^2(1)}{\sum_{i=1}^{d}\lambda_i} = \frac{\sum_{i=1}^{d}\lambda_i \chi_i^2(1)}{\mathrm{tr}(\boldsymbol{TV})},$$

where $\chi_i^2(f)$s are chi-squared distributed random variables with degrees of freedom $f$, and $\lambda_i$s are eigenvalues of $\boldsymbol{TV}$. Thus, $F_N^{CG}$ has the unit mean asymptotically (the proof given in Appendix A.3).

We construct an approximately level $\alpha$ test by rejecting a hypothesis for $F_N > c_{N,\alpha}$, where $c_{N,\alpha}$ is the upper $\alpha \times 100$ percentile of the null distribution. Since the critical value $c_{N,\alpha}$ involves unknown quantities, $\lambda_i$s, we propose two methods to calibrate $c_{N,\alpha}$:

**Simulation method for $c_{N,\alpha}$:**

We propose a simulation-based method to calibrate $c_{N,\alpha}$; a similar method was suggested by Brunner et al.[33] for data without censoring. This method replaces $\lambda_i$ with $\hat{\lambda}_i$, the eigenvalue of $\boldsymbol{T}\hat{\boldsymbol{V}}^{CG}$. Then, one generates random numbers $\chi_{i,(r)}^2(1)$, $i = 1, \ldots, d$, $r = 1, \ldots, R$ for the large number $R$ (e.g., $R = 1000$). Then, one obtains $c_{N,\alpha}$ as the upper $\alpha \times 100$ percent point of

$$F_{N,(r)}^{CG} = \frac{\sum_{i=1}^{d}\hat{\lambda}_i \chi_{i,(r)}^2(1)}{\mathrm{tr}(\boldsymbol{T}\hat{\boldsymbol{V}}^{CG})}, \qquad r = 1, \ldots, R.$$

**Analytical method for $c_{N,\alpha}$:**

We propose an analytical method to calibrate $c_{N,\alpha}$; a similar method was proposed for data without censoring[33,39]. The idea is to approximate the weighted chi-squared distribution by a chi-square distribution, $c\chi^2(f)$, where $c$ and $f$ are chosen to match the 1st and 2nd moments[49]. This leads to

$$c_{N,\alpha} = \frac{\chi^2(1-\alpha,\hat{f})}{\hat{f}}, \quad \hat{f} = \frac{\text{tr}^2(T\hat{V}^{CG})}{\text{tr}(T\hat{V}^{CG}T\hat{V}^{CG})}.$$

where $\chi^2(1-\alpha,\hat{f})$ is the upper $\alpha$ percentile of the chi-squared distribution with $\hat{f}$ degrees of freedom.

**3.4 Restricted follow-up**

A technical correction is necessary when survival data are collected within a restricted follow-up time. In this situation, the pairwise effects $w_{i\ell} = P(T_i > T_\ell) = -\int S_i^{\pm}(t) dS_\ell(t)$ may not be identifiable from data since the survival function $S_i(t)$ is not identifiable for some $t$ and some $i$. What we can identify from survival data is $S_i(t)$ on $t \in [0,\tau]$, where $\tau > 0$ is a number satisfying Assumption 2 in Appendix A3, namely,

$$\tau < \min_i [\sup\{u: G_i(u)S_i(u) > 0\}].$$

As in Dobler and Pauly[7], we modify the pairwise effect by

$$w_{i\ell} = P[\min(T_i,\tau) > \min(T_\ell,\tau)] + \frac{1}{2}P[\min(T_i,\tau) = \min(T_\ell,\tau)]$$
$$= -\int_0^\tau S_i^{\pm}(t) dS_\ell(t) + \frac{S_i(\tau)S_\ell(\tau)}{2}.$$

The last term in the right-hand side corrects the reduced amount of the pairwise effect due to the truncated integral. Accordingly, the estimator is also modified as

$$\widehat{w}_{i\ell}^{CG} = -\int_0^\tau \hat{S}_i^{CG\pm}(t) d\hat{S}_\ell^{CG}(t) + \frac{\hat{S}_i^{CG}(\tau)\hat{S}_\ell^{CG}(\tau)}{2}.$$

## 4. Software implementation

We implemented the proposed method in an R function surv.factorial(.) available in the R package *compound.Cox*[32]. The package includes other useful functions, CG.Clayton(.), CG.Frank(.), and CG.Gumbel(.), which can compute the CG estimators (Section 3.1).

The function surv.factorial(.) can compute the estimates of treatment effects $\hat{\boldsymbol{p}}^{CG}$, their SE and 95%CI (Section 3.2). It also computes the *F*-statistic and its critical value for testing $H_0: \boldsymbol{Cp} = \boldsymbol{0}$ under factorial designs (Section 3.3). The contrast matrix $\boldsymbol{C}$ and the follow-up end $\tau$ can be flexibly specified by users. For more details, the readers is referred to the package manual. We have checked the reliability of the R function by extensive simulation studies (Section 5).

For implementation, we assume that the copula is identical across treatments, namely, $C_{1\theta_1} = \cdots = C_{d\theta_d}$ and $\theta_1 = \cdots = \theta_d$. This is because the specifications of $d$ different copulas make the method more difficult to be used and interpreted by users.

## 5. Simulation

Simulation studies were conducted to investigate the performance of the proposed methods.

### 5.1 Simulation designs

The sample size was $n_i = 50$, or 100. We generated $T_{ij}$ and $U_{ij}$ from the Clayton copula model with $\theta_i = 2$ (Kendall's tau = 0.5) whose marginals follow exponential distributions,
$$T_{ij} \sim S_i(t) = \exp(-\lambda_i t), \qquad U_{ij} \sim G_i(t) = \exp(-\mu_i t),$$
for $i = 1, 2, \ldots, d$, $j = 1, \ldots, n_i$. We redefined $U_{ij}$ by $U_{ij} = \min(U_{ij}, \tau)$, where $\tau = 1$ is the follow-up end. We then obtained the censored data $\{(X_{ij}, \delta_{ij}); i = 1, \ldots, d, j = 1, \ldots, n_i\}$, where $X_{ij} = \min(T_{ij}, U_{ij})$ and $\delta_{ij} = 1\{T_{ij} \leq U_{ij}\}$.

Based on the data, we calculated the estimator $\hat{p}_i^{CG}$ and the 95%CI to see the accuracy of estimating $p_i$. We considered $d = 3$ for the one-way layout (3 levels), and $d = 6$ for the two-way layout ($2 \times 3 = 6$ levels). For the one-way layout, we calculated the *F*-statistic, $F_N^{CG}$, for testing
$$H_0: p_1 = p_2 = p_3 \quad \text{vs.} \quad H_1: p_1 \neq p_2, \text{ or } p_1 \neq p_3, \text{ or } p_2 \neq p_3$$
by setting the contrast matrix

$$C = I_3 - \frac{\mathbf{1}_3 \mathbf{1}_3'}{3} = \frac{1}{3}\begin{pmatrix} 1 & -1 & -1 \\ -1 & 1 & -1 \\ -1 & -1 & 1 \end{pmatrix}.$$

For the two-way layout ($2 \times 3 = 6$ levels), the true treatment effects $\boldsymbol{p}$ are

$$\boldsymbol{p} = (p_1, p_2, p_3, p_4, p_5, p_6)' = (\underbrace{p_{11}, p_{12}, p_{13}}_{\text{1st level in Factor A}}, \underbrace{p_{21}, p_{22}, p_{23}}_{\text{2nd level in Factor A}})'.$$

To test the null effect of Factor A, we calculated the $F$-statistic, $F_N^{CG}$, for no main effect in Factor A,

$$H_0: p_{11} + p_{12} + p_{13} = p_{21} + p_{22} + p_{23} \quad \text{vs.} \quad H_1: p_{11} + p_{12} + p_{13} \neq p_{21} + p_{22} + p_{23},$$

by setting

$$C = P_2 \otimes \frac{\mathbf{1}_3'}{3} = \frac{1}{6}\begin{pmatrix} 1 & 1 & 1 & -1 & -1 & -1 \\ -1 & -1 & -1 & 1 & 1 & 1 \end{pmatrix}.$$

The parameters $(\lambda_i, \mu_i)$ were set to yield realistic amount of treatment effects and censoring percentage $P(U_{ij} < T_{ij}) \times 100$. We made nine scenarios (Scenarios 1-6 for the one-way layout; Scenarios 7-9 for the two-way layout) for parameter settings (Table 1). The null hypothesis $H_0$ holds for Scenarios 1, 2, 7, and 8 while the alternative hypothesis $H_1$ holds for Scenarios 3, 4, 5, 6 and 9.

Based on 500 replications of simulated data, we assessed the accuracy of the proposed estimators for treatment effects $\boldsymbol{p}$ and proposed tests for $H_0$.

**Table 1**: Parameter settings for the simulation studies defined by nine scenarios. True treatment effects are beneficial (blue) or harmful (red) relative to the overall effect (black).

|  | Survival model: $S_i(t) = \exp(-\lambda_i t)$ | Censoring model: $G_i(t) = \exp(-\mu_i t)$, | True treatment effects |
|---|---|---|---|
| One-way | $(\lambda_1, \lambda_2, \lambda_3)$ | $(\mu_1, \mu_2, \mu_3)$ | $(p_1, p_2, p_3)$ |
| Scenario 1 | (1, 1, 1) | (1, 1, 1) | (0.5, 0.5, 0.5) |
| Scenario 2 | (1, 1, 1) | (1, 1.25, 1.5) | (0.5, 0.5, 0.5) |
| Scenario 3 | (1, 1.25, 1.5) | (1, 1, 1) | (0.547, 0.498, 0.455) |
| Scenario 4 | (1, 1.25, 1.5) | (1, 1.25, 1.5) | (0.547, 0.498, 0.455) |
| Scenario 5 | (1.25, 1, 0.75) | (1, 1, 1) | (0.447, 0.497, 0.556) |
| Scenario 6 | (1.25, 1, 0.75) | (1, 1.25, 1.5) | (0.447, 0.497, 0.556) |
| Two-way | $(\lambda_{11}, \lambda_{12}, \lambda_{13}, \lambda_{21}, \lambda_{22}, \lambda_{23})$ | $(\mu_{11}, \mu_{12}, \mu_{13}, \mu_{21}, \mu_{22}, \mu_{23})$ | $(p_{11}, p_{12}, p_{13}, p_{21}, p_{22}, p_{23})$ |
| Scenario 7 | (1, 1, 1, 1, 1, 1) | (1, 1, 1, 1, 1, 1) | (0.5, 0.5, 0.5, 0.5, 0.5, 0.5) |
| Scenario 8 | (1, 1.25, 1.5, 1, 1.25, 1.5) | (1, 1, 1, 1, 1, 1) | (0.55, 0.50, 0.46, 0.55, 0.50, 0.46) |
| Scenario 9 | (1, 1.25, 1.5, 1, 1, 1) | (1, 1.25, 1.5, 1, 1, 1) | (0.52, 0.47, 0.43, 0.52, 0.52, 0.52) |

## 5.2 Simulation results

We first show the results of the estimators and tests when the true value of $\theta_i = 2$ is known. Next, we turn out attention to the case, where $\theta_i = 2$ is possibly misspecified as $\theta_i = 1, 2, 3$, and 4.

When the true value $\theta_i = 2$ is known, the means of $\hat{p}_i^{CG}$ are close to the true value $p_i$, and hence, the estimators are nearly unbiased (Tables A1-A2 in Supplementary Materials). The variability (standard deviation, SD) of $\hat{p}_i^{CG}$ vanishes to zero when the sample size gets larger or the censored proportions get lower. This means that the estimators are consistent for the true values. The SDs are very close to the averages of the SEs, which implies the consistency of the jackknife estimates of the asymptotic variance. Consequently, the coverage probabilities of the 95%CI are close to the nominal value of 0.95. Overall, we observe the desirable performance for estimating $p_i$.

Table 2 shows the results for the proposed $F$-test when the true value of $\theta_i = 2$ is known. The type I error rates are well-controlled for the nominal level of $\alpha \in \{0.10, 0.05, 0.01\}$; see the rows of Scenarios 1, 2, 7, and 8. The proposed test has a reasonable amount of power; see the row of Scenarios 3, 4, 5, 6 and 9. The power gets higher when the sample size increases. There is no noticeable difference between the two methods for computing the critical values (Simulation method vs. Analytical method). Overall, the proposed test has desirable operating characteristics.

**Table 2**: Rejection rates for the proposed test based on 500 simulation runs under level $\alpha$.

| Model | Scenario | $n_i$ | Simulation method[†] | | | Analytical method[#] | | |
|---|---|---|---|---|---|---|---|---|
| | | | $\alpha = 0.10$ | $\alpha = 0.05$ | $\alpha = 0.01$ | $\alpha = 0.10$ | $\alpha = 0.05$ | $\alpha = 0.01$ |
| 1-way; $H_0$ | S1 | 50 | 0.108 | 0.062 | 0.018 | 0.106 | 0.060 | 0.018 |
| | | 100 | 0.076 | 0.040 | 0.010 | 0.082 | 0.038 | 0.012 |
| 1-way; $H_0$ | S2 | 50 | 0.102 | 0.062 | 0.018 | 0.100 | 0.060 | 0.016 |
| | | 100 | 0.080 | 0.046 | 0.012 | 0.076 | 0.046 | 0.008 |
| 1-way; $H_1$ | S3 | 50 | 0.348 | 0.242 | 0.080 | 0.350 | 0.244 | 0.080 |
| | | 100 | 0.512 | 0.402 | 0.164 | 0.510 | 0.396 | 0.154 |
| 1-way; $H_1$ | S4 | 50 | 0.304 | 0.204 | 0.068 | 0.312 | 0.198 | 0.066 |
| | | 100 | 0.470 | 0.346 | 0.142 | 0.466 | 0.342 | 0.132 |
| 1-way; $H_1$ | S5 | 50 | 0.404 | 0.306 | 0.138 | 0.402 | 0.306 | 0.140 |
| | | 100 | 0.644 | 0.522 | 0.284 | 0.636 | 0.510 | 0.280 |
| 1-way; $H_1$ | S6 | 50 | 0.396 | 0.264 | 0.144 | 0.386 | 0.260 | 0.138 |
| | | 100 | 0.558 | 0.434 | 0.224 | 0.552 | 0.428 | 0.222 |
| 2-way; $H_0$ | S7 | 50 | 0.084 | 0.052 | 0.012 | 0.084 | 0.054 | 0.006 |
| | | 100 | 0.104 | 0.050 | 0.014 | 0.106 | 0.048 | 0.016 |
| 2-way; $H_0$ | S8 | 50 | 0.092 | 0.050 | 0.014 | 0.090 | 0.044 | 0.012 |
| | | 100 | 0.104 | 0.048 | 0.010 | 0.102 | 0.050 | 0.010 |
| 2-way; $H_1$ | S9 | 50 | 0.342 | 0.218 | 0.078 | 0.334 | 0.214 | 0.074 |
| | | 100 | 0.488 | 0.356 | 0.184 | 0.488 | 0.350 | 0.180 |

[†] The critical value $c_{N,\alpha}$ is the upper $\alpha \times 100$ percent point of simulated samples (Section 3.3).
[#] The critical value is $c_{N,\alpha} = \chi^2_{\hat{f}, 1-\alpha}/\hat{f}$ with $\hat{f}$ degrees of freedom (Section 3.3).

Figures 1 (the box plots) shows the performance of the estimator $\hat{p}_i^{CG}$ under the misspecification of $\theta_i$ with $n_i = 100$. Under Scenario 1, the estimates are nearly unbiased and the coverage probabilities of 95%CI are close to 0.95 across all the misspecified values of $\theta_i$. This phenomenon is specific for Scenario 1, where the biases were cancelled out by equal censoring percentages across groups. With unequal censoring percentages as in Scenario 2, the estimates get biased as the specified value of $\theta_i$ deviates from the true value $\theta_i = 2$. Scenario 3 also yields some biases caused by misspecified value of $\theta_i$. However, even when $\theta_i$ is misspecified and the estimator $\hat{p}_i^{CG}$ is biased, the coverage probabilities remain close to 0.95 in many cases. However, for the misspecified value of $\theta_i = 0$, the obvious under-coverage is found.

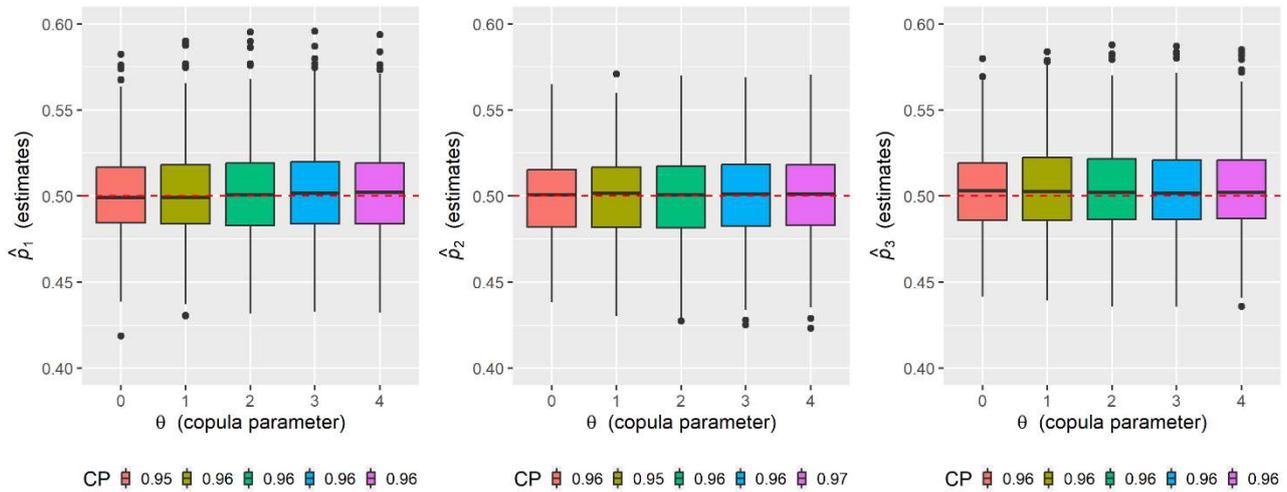

(a) **Scenario 1**: One-way layout with $p_1 = p_2 = p_3 = 0.500$

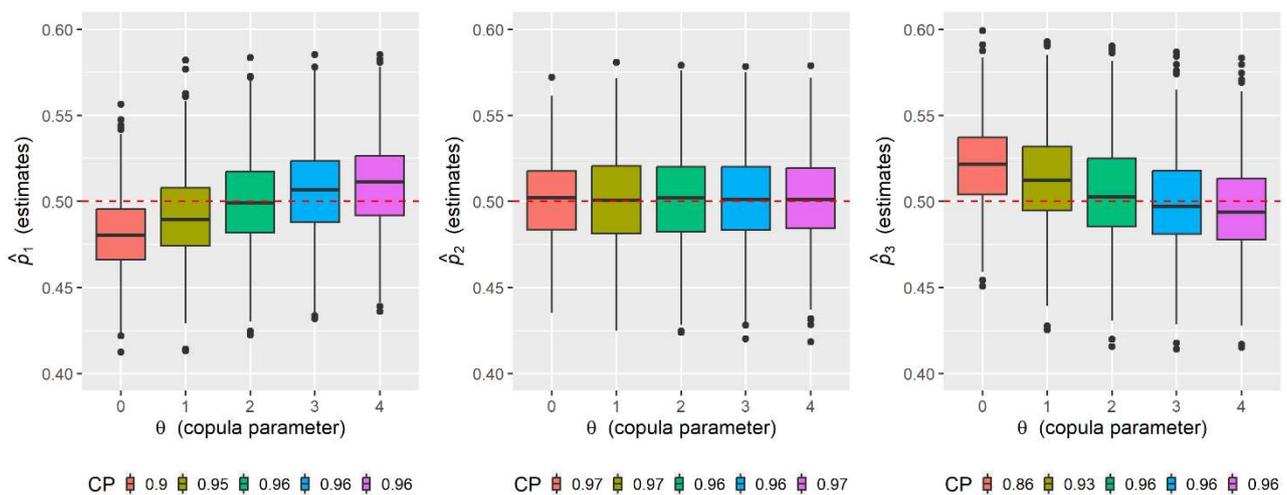

(b) **Scenario 2**: One-way layout with $p_1 = p_2 = p_3 = 0.500$

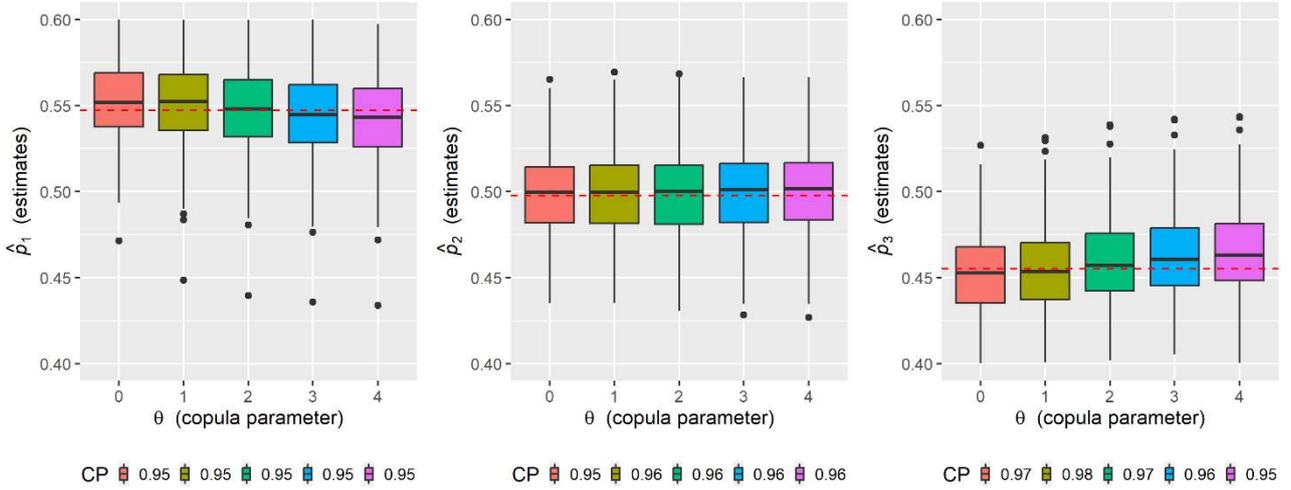

(c) **Scenario 3**: One-way layout with $p_1 = 0.552$, $p_2 = 0.497$, and $p_3 = 0.452$

**Figure 1**: Simulation results for $\hat{p}_i^{CG}$, $i = 1,2,3$, when the true value $\theta_i = 2$ is misspecified as $\theta_i = 0, 1, 2, 3, 4$. CP is to the coverage probability of the 95%CI.

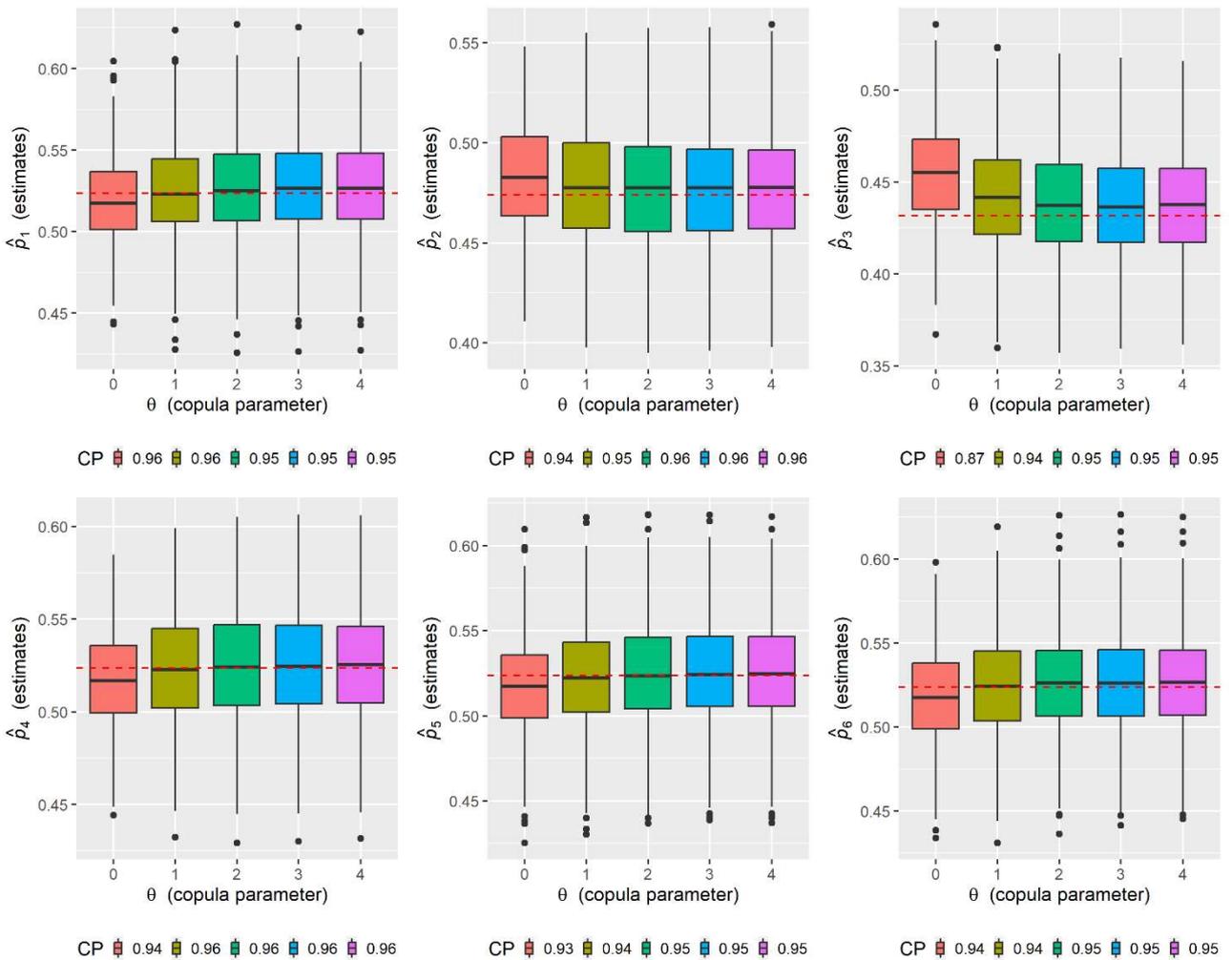

**Figure 2**: Simulation results for $\hat{p}_i^{CG}$, $i = 1,2,3$, when the true value $\theta_i = 2$ is misspecified as $\theta_i = 0, 1, 2, 3, 4$. A two-way layout is considered under Scenario 9 with $p_1 = 0.52$, $p_2 = 0.47$, $p_3 = 0.43$, and $p_4 = p_5 = p_6 = 0.52$. CP is to the coverage probability of the 95%CI.

Therefore, if the independent censoring assumption is mistakenly imposed, the estimator $\hat{p}_i^{CG}$ could be obviously biased and the CI may fail to cover the true value. Similar results are observed under Scenarios 4-6, which are given in Figure B1 of Supplementary Materials. These conclusions for the one-way layout continue to hold for the two-way layout, as seen for Scenarios 7-8 (Figures B2 of Supplementary Materials) and Scenario 9 (Figure 2).

## 6. Data analysis

We analyze breast cancer data available from an R Bioconductor package, *curatedBreastData*[50]. The data contain survival outcomes and other clinical covariates on 2719 breast cancer patients with advanced breast cancer. In the following analysis, we chose a subset of the data having the complete information on both disease-free survival (DFS) outcomes and treatment types. There are three treatment types: adjuvant, neoadjuvant, and mixed (both neoadjuvant and adjuvant). Excluding patients with missing DFS or treatment, we shall analyze a subset of 635 patients. We analyze this subset to compare three treatments in the one-way layout ($n_1 = 136$ adjuvant, $n_2 = 107$ mixed, and $n_3 = 392$ neoadjuvant patients).

Figure 3 displays estimated survival probabilities for DFS for the three treatment groups (adjuvant, neoadjuvant, or mixed treatment) by using the CG estimators with the Clayton copula. The adjuvant treatment yields higher survival probabilities than the mixed and neoadjuvant groups, for all the copula parameters, $\theta_i = 0$, $\theta_i = 2$, $\theta_i = 4$, and $\theta_i = 8$. Another obvious pattern is that estimated survival probabilities reduce remarkably when the copula parameter moves from $\theta_i = 0$ to $\theta_i = 8$. This demonstrates the strong impact of dependent censoring on estimated survival.

To check the significance of the difference of treatments, we shall perform a hypothesis test. The logrank test rejected the global hypothesis $H_0^S: S_1(t) = S_2(t) = S_3(t) \ \forall t$ in favor of survival difference (chi-squared statistic=17.5, critical value=5.99, P-value<0.05). We stress that this conclusion was derived under the independent censoring assumption, which corresponds to survival difference at the parameter $\theta_i = 0$ (see the upper left panel of Figure 3).

In order to see how the conclusion might change under dependent censoring, we applied the proposed $F$-test. To this end, we perform the test under the parameters $\theta_i = 0$, $\theta_i = 2$, $\theta_i = 4$, and $\theta_i = 8$. Table 3 shows that the $F$-test rejected $H_0: p_1 = p_2 = p_3 = 1/2$ at 5% level for all the values of $\theta_i$. Hence, even if the independent censoring assumption violates, the hypothesis $H_0^S: S_1(t) = S_2(t) = S_3(t) \ \forall t$ is rejected at 5% level. This strengthens the conclusion from the logrank tests by relaxing the independent censoring assumption.

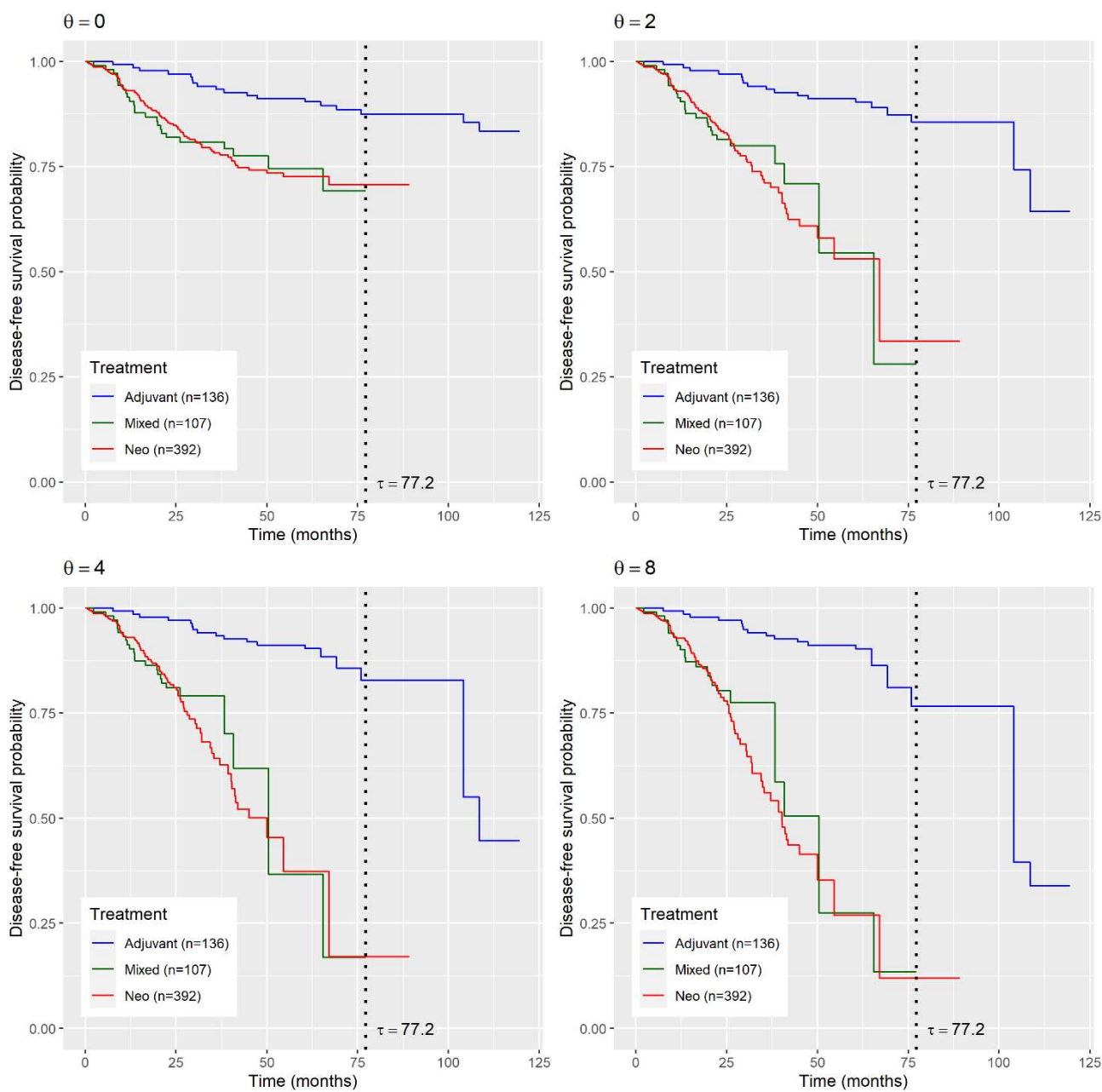

**Figure 3**: Copula-Graphic estimates of DFS probabilities for the breast cancer data based on three treatments: adjuvant, neoadjuvant, and mixed (both neoadjuvant and adjuvant). The four panels correspond to the fitted Clayton copula with the parameters $\theta_i = 0$, $\theta_i = 2$, $\theta_i = 4$, and $\theta_i = 8$.

Table 3: Testing the equality of treatment effects on DFS based on the breast cancer data.

| Copula parameter | F-value $F_N^{CG}$ | Critical value (5%) $c_{N,\alpha}$: simulation | Critical value (5%) $c_{N,\alpha}$: analytical | P-value |
|---|---|---|---|---|
| $\theta = 0$ (Kendall's tau=0.00) | 133.979 | 3.073 | 3.044 | <0.0001 |
| $\theta = 2$ (Kendall's tau=0.50) | 15.453 | 3.763 | 3.515 | 0.0009 |
| $\theta = 4$ (Kendall's tau=0.67) | 22.764 | 3.602 | 3.405 | 0.0001 |
| $\theta = 8$ (Kendall's tau=0.80) | 39.807 | 3.364 | 3.250 | <0.0001 |

As we detected significant differences in treatment effects, we estimated them by the proposed estimators. Table 4 shows the estimates of $(p_1, p_2, p_3)$ using the proposed estimator by fitting the Clayton copula with the parameter $\theta_i$. For instance, if $\theta_i = 8$ is assumed, the adjuvant treatment yields survival advantage ($\hat{p}_1^{CG} = 0.761 > 0.5$) over other treatments. The difference is significant since its 95%CI did not cover 0.5 (Table 4). Indeed, for any parameter, the adjuvant treatment yields the best and significant survival advantage. This extremely confirms the advantage of the adjuvant treatment under dependent censoring. The difference between the neoadjuvant and adjuvant treatments depends on $\theta_i$ and the comparison is hence inconclusive.

**Table 4**: Estimation of treatment effects $p_i$ on DFS based on the breast cancer data.

| Copula parameter | Treatment | Parameter | Estimate | SE | 95%CI |
|---|---|---|---|---|---|
| $\theta = 0$ (Kendall's tau=0.00) | Adjuvant | $p_1$ | 0.664 | 0.014 | (0.638, 0.691) |
| | Mixed | $p_2$ | 0.283 | 0.013 | (0.258, 0.307) |
| | Neoadjuvant | $p_3$ | 0.554 | 0.015 | (0.524, 0.583) |
| $\theta = 2$ (Kendall's tau=0.50) | Adjuvant | $p_1$ | 0.725 | 0.034 | (0.659, 0.792) |
| | Mixed | $p_2$ | 0.341 | 0.032 | (0.279, 0.404) |
| | Neoadjuvant | $p_3$ | 0.436 | 0.055 | (0.329, 0.544) |
| $\theta = 4$ (Kendall's tau=0.67) | Adjuvant | $p_1$ | 0.754 | 0.034 | (0.686, 0.821) |
| | Mixed | $p_2$ | 0.364 | 0.027 | (0.310, 0.417) |
| | Neoadjuvant | $p_3$ | 0.389 | 0.048 | (0.259, 0.482) |
| $\theta = 8$ (Kendall's tau=0.80) | Adjuvant | $p_1$ | 0.761 | 0.028 | (0.706, 0.815) |
| | Mixed | $p_2$ | 0.376 | 0.023 | (0.332, 0.421) |
| | Neoadjuvant | $p_3$ | 0.369 | 0.035 | (0.300, 0.438) |

## 7. Conclusion and discussion

We develop a novel method for factorial survival analysis under a copula-based dependent censoring model. While the majority of the traditional survival analyses were formulated under the independent censoring model, we try to challenge this assumption in a situation of factorial designs. This article is the first attempt to examine dependent censoring in factorial designs. Even though copula-based dependent censoring models have already been considered in a variety of settings and applications[12,14,16,17,21,22,27-31], we extend these works toward a new setting: factorial designs.

While factorial designs refer to experimentally controlled designs (e.g., randomized clinical trials), the proposed method can also be applicable for observationally collected data with treatment groups as covariates. Indeed, we analyze survival data of 635 breast cancer patients treated by one of three treatments (adjuvant, neoadjuvant, or mixed). The proposed method offers a nonparametric means of estimating treatment effects without assuming any model, such as the proportional hazards model.

Significant survival advantage of the adjuvant treatment is found over the neoadjuvant and mixed treatments. This conclusion is shown to be robust for a variety of dependent censoring scenarios. Nonetheless, the estimated advantage of the adjuvant treatment may be biased since the treatments were not experimentally controlled (i.e., not randomized) and the data were collected observationally. For instance, if the neoadjuvant treatment were administered mainly for patients with poor survival prognosis, their survival would be inherently poor, irrespective of the effect of treatments. Therefore, we cannot exclude the possible bias of the estimated treatment effects due to unbalanced patient characteristics across treatments. It is of interest to develop a method to utilize covariates to adjust for potential biases.

Our simulations showed that the proposed *F*-test properly controls type I error rates and yields reasonable power of detection. We observed that the two methods to calibrate the critical values of the *F*-statistic exhibited similar performance. One method tries to approximate the null distribution by the Monte Carlo method while the other method by the best-fitting chi-square distribution. The Monte Carlo method has the asymptotic validity, yet it has some random fluctuation even when 1000 repetitions are used. The chi-square approximation is asymptotically valid only for its first and second moments, yet its analytical simplicity makes it a recommended choice.

In our numerical analyses, the Clayton copula was fitted. Nonetheless, other copulas could be tried, such as the Frank copula, especially when negative dependence is suspected for dependent censoring[12]. Our R package (Section 4) allows users to choose the Gumbel and Frank copulas, though we reported our results only for the Clayton copula due to the space limitation. Indeed, it is difficult to justify a suitable copula function by survival data.

The metric of treatment effects is formulated on the basis of pairwise comparison, which is free from model assumptions, such as the proportional hazards model. Recently, pairwise comparison has been adopted for many clinical trials with multiple endpoints[51] with aid of Buyse's generalized pairwise comparison[52]. This approach was extended to survival outcomes with independent censoring by Péron, et al.[53] along with the asymptotic theory[54]. Extension of the generalized pairwise comparison under dependent censoring is of great interest; see a relevant work for the extension to competing risks outcomes[55].

**Acknowledgement**

Emura T was supported financially by JSPS KAKENHI (22K11948). Ditzhaus M was funded by the Deutsche Forschungsgemeinschaft (grant no. DI 2906/1-2). A draft of this manuscript was presented by Emura T in an organized session (EO350: Beyond proportional hazards and standard survival), CM Statistics 2022, London. Comments from the audience helped improve the article.

# Appendices

## A1. Examples of the copulas

While there are a large number of copulas[40,41], we pick up a few examples. Besides the Clayton copula introduced in the main text, we provide three copulas below:

**Example 1: The Gumbel copula (Gumbel[56]):**

$$C_{i\theta_i}(u,v) = \exp\left[-\{(-\log u)^{\theta_i+1} + (-\log v)^{\theta_i+1}\}^{\frac{1}{\theta_i+1}}\right], \quad \theta_i \geq 0.$$

The copula gives Kendall's tau $\tau_{\theta_i} = \theta_i/(\theta_i + 1)$, producing positive dependence ($\theta_i > 0$) and independence ($\theta_i = 0$). The Gumbel copula was applied to a number of survival data, such as breast cancer data for modeling dependence between overall survival and time-to-relapse[57].

**Example 2: The Frank copula (Frank[58]):**

$$C_{i\theta_i}(u,v) = -\frac{1}{\theta_i}\log\left\{1 + \frac{(e^{-\theta_i u} - 1)(e^{-\theta_i v} - 1)}{e^{-\theta_i} - 1}\right\}, \quad \theta_i \in (-\infty, 0) \cup (0, \infty).$$

The Frank copula yields Kendall's tau

$$\tau_{\theta_i} = 1 - \frac{4}{\theta_i}\left(1 - \frac{1}{\theta_i}\int_0^{\theta_i} \frac{t}{e^t - 1} dt\right)$$

The Frank copula produces negative dependence ($\theta_i < 0$) and positive dependence ($\theta_i > 0$). The Frank copula converges to the independence copula when $\theta_i \to 0$. The Frank copula was applied to a number of survival data. For instance, the dependence structure of the disease-free survival and overall survival in patients with cervical cancer is best modeled by the Frank copula compared to many other copulas[59]. Huang et al.[60] also chose the Frank model for modeling dependence between death and readmission in patients with colorectal cancer.

**Example 3: The Farlie-Gumbel-Morgensterm (FGM) copula (Morgenstern[61]):**

$$C_{i\theta_i}(u,v) = uv\{1 + \theta_i(1-u)(1-v)\}, \quad -1 \leq \theta_i \leq 1.$$

The FGM copula produces negative dependence ($\theta_i < 0$), positive dependence ($\theta_i > 0$), and independence ($\theta_i = 0$), having Kendall's tau $\tau_{\theta_i} = 2\theta_i/9$. The FGM copula was used in applications to survival data[59], reliability data[62], chemical data[63], sports data[64], and econometric data [68].

## A2. Examples of the CG estimator

Under the Gumbel copula generated by $\phi_{i\theta_i}(t) = (-\log t)^{\theta_i+1}$, one has the CG estimator

$$\hat{S}_i^{CG}(t) = \exp\left[-\left(\sum_{j:\,X_{ij}\leq t,\,\delta_{ij}=1}\left[\left\{-\log\left(\frac{\bar{Y}_i(X_{ij})-1}{n_i}\right)\right\}^{\theta_i+1} - \left\{-\log\left(\frac{\bar{Y}_i(X_{ij})}{n_i}\right)\right\}^{\theta_i+1}\right]\right)^{\frac{1}{\theta_i+1}}\right].$$

Similarly, one can obtain the CG estimator under the Frank copula generated by $\phi_{i\theta_i}(t) = -\log\{(e^{-\theta_i t}-1)/(e^{-\theta_i}-1)\}$. The estimator is given by

$$\hat{S}_i^{CG}(t) = -\frac{1}{\theta_i}\log\left[1+\exp\left(-\sum_{j:\,X_{ij}\leq t,\,\delta_{ij}=1}\left\{\exp\left(-\theta_i\frac{\bar{Y}_i(X_{ij})-1}{n_i}\right) - \exp\left(-\theta_i\frac{\bar{Y}_i(X_{ij})}{n_i}\right)\right\}\right)\right].$$

## A3: Assumptions and proofs for asymptotic normality

To formulate the asymptotic distributions of the proposed estimator and the *F*-test, we impose some assumptions. Let $\boldsymbol{n} = (n_1, \ldots, n_d)'$ be a vector of sample sizes, and $N = \sum_{i=1}^{d} n_i$ be the total sample size. We impose the following assumptions:

***Assumption 1***: *As* $N \to \infty$,

$$\frac{\boldsymbol{n}'}{N} = \frac{(n_1,\,n_2,\ldots,n_d)}{N} \to \boldsymbol{\lambda} = (\lambda_1,\,\lambda_2,\ldots,\lambda_d),$$

*for* $\lambda_i \in (0,1)$, $i = 1, 2, \ldots, d$.

***Assumption 2***: *There exists* $\tau > 0$ *such that* $\pi(\tau) = P(X_{ij} \geq \tau) > 0$.

***Assumption 3***: $\phi'_{i\theta_i}(t) = d\phi_{i\theta_i}(t)/dt$ *and* $\psi'_{i\theta_i}(t) = d\psi_{i\theta_i}(t)/dt$ *are bounded on* $(\pi(\tau), 1)$, *where* $\psi_{i\theta_i}(t) = -u\phi'_{i\theta_i}(t)$.

Assumption 1 avoids extremely unbalanced sample sizes across treatments[7,39]. This assumption is widely employed for the asymptotic theory with multi-group comparative analyses not restricted to survival analysis (e.g., Ditzhaus and Smaga[65]).

Assumption 2 ensures that the estimator $\hat{S}_i^{CG}(.)$ is uniformly consistent for $S_i(.)$ on $[0, \tau)$ as long as the model (4) is correct: see Theorem 1 of Rivest and Wells[30].

Assumption 3 is a mild regularity condition of copulas, which holds for the Clayton, Gumbel, and Frank copulas. For the Gumbel copula, it follows that $\phi'_{i\theta_i}(t) = (\theta_i + 1)(-\log t)^{\theta_i}/t$ and $\psi'_{i\theta_i}(t) = -\theta_i(\theta_i + 1)(-\log t)^{\theta_i - 1}/t$ are both bounded in $(\varepsilon, 1)$ for $\varepsilon > 0$ with the maxima attained at $t = \varepsilon$. The cases of the Clayton and Frank copulas follows similarly.

We first establish the asymptotic normality of $\hat{\mathbf{S}}^{CG}(.) = (\hat{S}_1^{CG}(.), \ldots, \hat{S}_d^{CG}(.))'$, which is an extension of the results for $d = 1$ by Rivest and Wells[30] and $d = 2$ by Emura and Hsu[31]. Let $\mathbf{diag}(\lambda)$ be a diagonal matrix with the diagonal elements $\lambda$. Then, as $N \to \infty$,

$$\sqrt{N}\left(\hat{\mathbf{S}}^{CG}(t) - \mathbf{S}(t)\right)$$

$$= \begin{pmatrix} \sqrt{n_1/N} & \cdots & 0 \\ \vdots & \ddots & \vdots \\ 0 & \cdots & \sqrt{n_d/N} \end{pmatrix}^{-1} \begin{pmatrix} \sqrt{n_1}\left(\hat{S}_1^{CG}(t_1) - S_1(t_1)\right) \\ \vdots \\ \sqrt{n_d}\left(\hat{S}_d^{CG}(t_d) - S_d(t_d)\right) \end{pmatrix} \xrightarrow{D} \{\mathbf{diag}(\lambda)\}^{-1/2}\mathbf{U},$$

where $\mathbf{U} = (U_1, \ldots, U_d)'$ are mutually independent Gaussian processes whose covariance $\Gamma_i(s, t) = \text{Cov}(U_i(s), U_i(t))$ is defined in Rivest and Wells[30]. The convergence $\xrightarrow{D}$ means a weak convergence in a metric space $\{D[0, \tau]\}^d$, where $D[0, \tau]$ is the space of *cadlag* functions on $[0, \tau]$ (i.e., functions that are right-continuous with left-hand limits: see Chapter 3 of Billingsley[66]).

Next, by applying the functional delta method[67] to the Hadamard differentiable functional $(S_i, S_\ell) \mapsto -\int S_i^\pm(t)dS_\ell(t)$, one has the asymptotic linear expression:

$$\sqrt{N}(\hat{w}_{ij}^{CG} - w_{ij}) = \sqrt{N}\left(-\int \hat{S}_i^{CG\pm}(t)d\hat{S}_j^{CG}(t) + \int S_i^\pm(t)dS_j(t)\right)$$

$$= \sqrt{\frac{N}{n_i}}\int \sqrt{n_i}(\hat{S}_i^{CG\pm} - S_i^\pm)(t)dS_j(t) + \sqrt{\frac{N}{n_j}}\int \sqrt{n_j}(\hat{S}_j^{CG\pm} - S_j^\pm)(t)dS_i(t) + o_P(1).$$

This leads to

$$\sqrt{N}(\hat{w}_{ij}^{CG} - w_{ij}) \xrightarrow{D} \lambda_i^{-1/2}\int U_i^\pm(t)dS_j(t) + \lambda_j^{-1/2}\int U_j^\pm(t)dS_i(t),$$

where the convergence $D$ means the convergence in distribution. Thus, one has the asymptotic normality of $\hat{\mathbf{w}}^{CG} = (\hat{w}_{11}^{CG}, \ldots, \hat{w}_{1d}^{CG} : \cdots : \hat{w}_{d1}^{CG}, \ldots, \hat{w}_{dd}^{CG})'$, namely,

$$\sqrt{N}(\hat{\mathbf{w}}^{CG} - \mathbf{w}) \xrightarrow{D} \mathbf{W} = N_{d^2}(\mathbf{0}, \mathbf{\Omega}),$$

where $\mathbf{\Omega}$ is a $d^2 \times d^2$ covariance matrix with the element

$$\Omega_{ij,k\ell} = \mathrm{Cov}(W_{ij}, W_{k\ell}) \doteq \mathrm{Cov}\left(-\int \hat{S}_i^{\mathrm{CG}\pm}(t) d\hat{S}_j^{\mathrm{CG}}(t), -\int \hat{S}_k^{\mathrm{CG}\pm}(t) d\hat{S}_\ell^{\mathrm{CG}}(t)\right),$$

For $i = j$ or $k = l$, one has the degenerated case $\Omega_{ij,kl} = 0$. Also, for $i \neq j$, $k \neq l$, $i \neq k$, and for $j \neq l$, one has the independent case $\Omega_{ij,kl} = 0$. For non-degenerate and non-independent cases,

$$\Omega_{ij,kl} = \mathrm{Cov}(W_{ij}, W_{kl})$$

$$= \begin{cases} \dfrac{1}{\lambda_j}\iint \Gamma_j(s,t)dS_i(s)dS_i(t) + \dfrac{1}{\lambda_i}\iint \Gamma_i(s,t)dS_j(s)dS_j(t) & \text{for } i \neq j, k \neq l, i = k, \text{and } j = l, \\ \dfrac{1}{\lambda_i}\iint \Gamma_i(s,t)dS_j(s)dS_l(t) & \text{for } i \neq j, k \neq l, i = k, \text{and } j \neq l, \\ \dfrac{1}{\lambda_j}\iint \Gamma_j(s,t)dS_i(s)dS_k(t) & \text{for } i \neq j, k \neq l, i \neq k, \text{and } j = l. \end{cases}$$

It follows that $\sqrt{N}A(\hat{w}^{\mathrm{CG}} - w) \xrightarrow{D} AW$, which is equivalent to

$$\sqrt{N}(\hat{p}^{\mathrm{CG}} - p) \xrightarrow{D} N_d(\mathbf{0}, V),$$

where $V = A\Omega A'$. Finally, under the null hypothesis, one has

$$\sqrt{N}T\hat{p}^{\mathrm{CG}} \xrightarrow{d} N_d(\mathbf{0}, TVT') = (TVT')^{1/2} N_d(\mathbf{0}, I_d).$$

By $TT = T = T'$ and the well-known results from quadratic forms,

$$N\hat{p}^{\mathrm{CG}\prime}T\hat{p}^{\mathrm{CG}} = \left(\sqrt{N}T\hat{p}^{\mathrm{CG}}\right)'\left(\sqrt{N}T\hat{p}^{\mathrm{CG}}\right) \xrightarrow{d} N_d(\mathbf{0}, I_d)'(TVT')N_d(\mathbf{0}, I_d) = \sum_{i=1}^{d} \lambda_i \chi_i^2(1).$$

Finally, we have

$$F_N^{\mathrm{CG}} = \frac{N\hat{p}^{\mathrm{CG}\prime}T\hat{p}^{\mathrm{CG}}}{\mathrm{tr}(T\hat{V}^{\mathrm{CG}})} = \frac{N\hat{p}^{\mathrm{CG}\prime}T\hat{p}^{\mathrm{CG}}}{\mathrm{tr}(TV)} \times \frac{\mathrm{tr}(TV)}{\mathrm{tr}(T\hat{V}^{\mathrm{CG}})} \xrightarrow{d} \frac{\sum_{i=1}^d \lambda_i \chi_i^2(1)}{\sum_{i=1}^d \lambda_i} \times 1.$$

## Supplementary data

Online supplementary materials related to this article are:
- additional simulation results
- the R code for the simulation studies
- the R code for the data analysis

**References**
[1] Montgomery DC. *Design and Analysis of Experiments*. John Wiley & Sons, 2019.
[2] Mishra P, Singh U, Pandey CM, Mishra P, Pandey G. Application of student's t-test, analysis of variance, and covariance. *Annals of Cardiac Anaesthesia*, 2019, **22**(4), 407.
[3] Futschik A, Taus T, Zehetmayer S. An omnibus test for the global null hypothesis. *Statistical Methods in Medical Research*, 2019, **28**(8), 2292-2304.
[4] Fizazi K, Foulon S, Carles J, Roubaud G, *et al*. Abiraterone plus prednisone added to androgen deprivation therapy and docetaxel in de novo metastatic castration-sensitive prostate cancer (PEACE-1): a multicentre, open-label, randomised, phase 3 study with a 2× 2 factorial design. *The Lancet*, 2022, **399**(10336), 1695-1707.
[5] Freidlin B, Korn E. Two-by-Two factorial cancer treatment trials: is sufficient attention being paid to possible interactions?. *JNCI: Journal of the National Cancer Institute*, 2017, **109**(9):djx146.
[6] Tewari KS, Sill MW, et al. Improved survival with bevacizumab in advanced cervical cancer. *New England Journal of Medicine*, 2014, **370**(8), 734-743.
[7] Dobler D, Pauly M. Factorial analyses of treatment effects under independent right-censoring. *Statistical Methods in Medical Research*, 2020, **29**(2): 325-343.
[8] Gorfine M, Schlesinger M. Hsu L. K-sample omnibus non-proportional hazards tests based on right-censored data. *Statistical Methods in Medical Research*, 2020, **29**, 2830–2850.
[9] Ditzhaus M, Dobler D, Pauly M. Inferring median survival differences in general factorial designs via permutation tests. *Statistical Methods in Medical Research*, 2021, **30**(3), 875-891.
[10] Ditzhaus M, Genuneit J, Janssen A, Pauly M. CASANOVA: Permutation inference in factorial survival designs. *Biometrics,* 2023, https://doi.org/10.1111/biom.13575.
[11] Klein JP, Moeschberger ML. *Survival Analysis: Techniques for Censored and Truncated Data.* New York: Springer, 2003.
[12] Braekers R, Veraverbeke N. A copula-graphic estimator for the conditional survival function under dependent censoring. *Canadian Journal of Statistics,* 2005*,* **33**(3), 429-447.
[13] Lu Z, Zhang W. Semiparametric likelihood estimation in survival models with informative censoring. *Journal of Multivariate Analysis*, 2012, **106**, 187-211.
[14] Emura T, Michimae H. A copula-based inference to piecewise exponential models under dependent censoring, with application to time to metamorphosis of salamander larvae, *Environ Ecol Stat*, 2017, **24**(1), 151-173.
[15] Staplin N, Kimber A, Collett D, Roderick P. Dependent censoring in piecewise exponential survival models. *Statistical Methods in Medical Research,* 2015, **24**(3), 325-341.
[16] Emura T, Chen YH. Gene selection for survival data under dependent censoring, a copula-based approach, *Statistical Methods in Medical Research*, 2016, **25**(6): 2840–57.
[17] Emura T, Chen YH. *Analysis of Survival Data with Dependent Censoring: Copula-Based Approaches*: Springer, 2018.
[18] Dettoni R, Marra G, Radice R. Generalized link-based additive survival models with informative censoring. *Journal of Computational and Graphical Statistics*, 2020, **29**(3), 503-512.
[19] Andersen PK, Perme MP. Pseudo-observations in survival analysis. *Statistical Methods in Medical Research*, 2010, **19**(1), 71-99.
[20] Collett D, *Modelling Survival Data in Medical Research, Third Edition*, Chapman & Hall, 2015.
[21] Xu J, Ma J, Connors MH, Brodaty H. Proportional hazard model estimation under dependent censoring using copulas and penalized likelihood. *Statistics in Medicine*, 2018, **37**(14), 2238-2251.
[22] Moradian H, Larocque D, Bellavance F. Survival forests for data with dependent censoring. *Statistical Methods in Medical Research,* 2019, **28**(2), 445-461.
[23] Schneider S, Demarqui FN, Colosimo EA, Mayrink VD. An approach to model clustered survival data with dependent censoring. *Biometrical Journal*, 2020, **62**(1), 157-174.
[24] Schneider S, Demarqui F, de Freitas Costa E. Free-ranging dogs' lifetime estimated by an approach for long-term survival data with dependent censoring. *Environ. Ecol. Stat.* 2022, **29**(4), 869-911.
[25] Deresa NW, Van Keilegom I. Flexible parametric model for survival data subject to dependent censoring. *Biometrical Journal*, 2020, **62**(1): 136-156
[26] Deresa NW, Van Keilegom I. A multivariate normal regression model for survival data subject to different types of dependent censoring. *Computational Statistics & Data Analysis*, 2020, **144**, 106879.

[27] Li D, Hu XJ, Wang R. Evaluating association between two event times with observations subject to informative censoring. *Journal of the American Statistical Association*, 2021, doi:10.1080/01621459.2021.1990766.

[28] Deresa NW, Van Keilegom I Antonio K. Copula-based inference for bivariate survival data with left truncation and dependent censoring. *Insurance: Mathematics and Economics*, 2022, **107**, 1-21.

[29] Czado C, Van Keilegom I. Dependent censoring based on parametric copulas. *Biometrika*, 2022, asac067.

[30] Rivest LP, Wells MT. A martingale approach to the copula-graphic estimator for the survival function under dependent censoring. *Journal of Multivariate Analysis,* 2001, **79**(1), 138-155.

[31] Emura T, Hsu JH. Estimation of the Mann–Whitney effect in the two-sample problem under dependent censoring. *Computational Statistics & Data Analysis*, 2020, **150**, 106990.

[32] Emura T, Matsui S, Chen HY. (2019). compound.Cox: univariate feature selection and compound covariate for predicting survival. *Computer Methods and Programs in Biomedicine, 168*, 21-37.

[33] Brunner E, Konietschke F, Pauly M, Puri ML. Rank-based procedures in factorial designs: hypotheses about non-parametric treatment effects. *Journal of the Royal Statistical Society: Series B*, 2017, **79**(5), 1463-1485.

[34] Efron B. The two sample problem with censored data. *In Proceedings of the Fifth Berkeley Symposium on Mathematical Statistics and Probability* (Vol. 4, No. University of California Press, Berkeley, CA, pp. 831-853), 1967.

[35] Koziol JA, Jia Z. The concordance index C and the Mann–Whitney parameter Pr (X> Y) with randomly censored data. *Biometrical Journal,* 2009, **51**(3), 467-474.

[36] Dobler D, Pauly M. Bootstrap-and permutation-based inference for the Mann–Whitney effect for right-censored and tied data. *Test,* 2018, **27**(3), 639-658.

[37] Dobler D, Friedrich S, Pauly M. Nonparametric MANOVA in meaningful effects. *Annals of the Institute of Statistical Mathematics*, 2020, **72**(4), 997-1022.

[38] Nowak CP, Mütze T, Konietschke F. Group sequential methods for the Mann-Whitney parameter. *Statistical Methods in Medical Research*, 2022, **31**(10) 2004-2020.

[39] Brunner E, Puri ML Nonparametric methods in factorial designs. *Statistical Papers,* 2001, **42**(1), 1-52.

[40] Nelsen RB. *An introduction to Copulas*: Springer Science & Business Media, 2006.

[41] Durante F, Sempi C. *Principles of Copula Theory*. Boca Raton, FL: CRC press, 2016.

[42] Clayton DG. A model for association in bivariate life tables and its application in epidemiological studies of familial tendency in chronic disease incidence. *Biometrika,* 1978, **65**(1), 141-151.

[43] Zheng M, Klein JP. Estimates of marginal survival for dependent competing risks based on an assumed copula. *Biometrika,* 1995, **82**(1), 127-138.

[44] Tsiatis A. A nonidentifiability aspect of the problem of competing risks. *Proc. Natn. Acad. Sci. USA*, 1975, **72**: 20-22.

[45] De Uña-Álvarez J, Veraverbeke N. Generalized copula-graphic estimator. *Test*, 2013, **22**(2), 343-360.

[46] De Uña-Álvarez J, Veraverbeke N. Copula-graphic estimation with left-truncated and right-censored data. *Statistics*, 2017, **51**(2), 387-403.

[47] Efron B, Tibshirani RJ. *An Introduction to the Bootstrap*. CRC press, 1994.

[48] Perme MP, Manevski D. Confidence intervals for the Mann–Whitney test. *Statistical Methods in Medical Research*, 2019, **28**(12), 3755-3768.

[49] Box GE. Some theorems on quadratic forms applied in the study of analysis of variance problems, I. Effect of inequality of variance in the one-way classification. *The Annals of Mathematical Statistics*, **25**(2), 1954, 290-302.

[50] Planey K. curatedBreastData: Curated breast cancer gene expression data with survival and treatment information. *CRAN*, 2022, R package version 2.26.0

[51] Deltuvaite‐Thomas V, Verbeeck J, Burzykowski T, Buyse M, et al. (2022). Generalized pairwise comparisons for censored data: An overview. *Biometrical Journal*, https://doi.org/10.1002/bimj.202100354.

[52] Buyse M. Generalized pairwise comparisons of prioritized outcomes in the two-sample problem. *Statistics in Medicine*, 2010, **29**(30), 3245-3257.

[53] Péron J, Buyse M, Ozenne B, Roche L, Roy P. An extension of generalized pairwise comparisons for prioritized outcomes in the presence of censoring. *Statistical Methods in Medical Research*, 2018, **27**(4), 1230-1239.

[54] Ozenne B, Budtz-Jørgensen E, Péron J. The asymptotic distribution of the Net Benefit estimator in presence of right-censoring. *Statistical Methods in Medical Research,* 2021, **30**(11), 2399-2412.

[55] Cantagallo E, De Backer M, Kicinski M, Ozenne B, Collette L, Legrand C. et al. A new measure of treatment effect in clinical trials involving competing risks based on generalized pairwise comparisons. *Biometrical Journal*, 2021, **63**(2), 272-288.

[56] Gumbel EJ. Distributions des valeurs extremes en plusiers dimensions. *Publ. Inst. Statist. Univ. Paris, 1960,* **9**, 171-173.

[57] Peng M, Xiang L. Joint regression analysis for survival data in the presence of two sets of semi-competing risks. *Biometrical Journal*, **61**(6), 2019, 1402-1416.

[58] Frank MJ. On the simultaneous associativity of F(x,y) and x + y – F(x,y). *Aequationes Math.*, 1979, **19**:194-226.

[59] de Oliveira Peres MV, Achcar JA, Martinez EZ. Bivariate lifetime models in presence of cure fraction: a comparative study with many different copula functions. *Heliyon*, 2020, **6**(6), e03961.

[60] Huang XW, Wang W, Emura T. A copula-based Markov chain model for serially dependent event times with a dependent terminal event. *Japanese Journal of Statistics and Data Science*, 2021, **4**(2), 917-951.

[61] Morgenstern D. Einfache beispiele zweidimensionaler verteilungen. *Mitteilingsblatt fur Mathematische Statistik,* 1956, **8**, 234-235.

[62] Ota S, Kimura M. Effective estimation algorithm for parameters of multivariate Farlie–Gumbel–Morgenstern copula. *Japanese Journal of Statistics and Data Science*, 2021, **4**(2), 1049-1078.

[63] Susam SO. A multi-parameter Generalized Farlie-Gumbel-Morgenstern bivariate copula family via Bernstein polynomial. *Hacettepe Journal of Mathematics and Statistics*, 2022, **51**(2): 618 – 631.

[64] de Oliveira RP, de Oliveira Peres MV, dos Santos MR, Martinez EZ, Achcar JA. A Bayesian inference approach for bivariate Weibull distributions derived from Roy and Morgenstern methods. *Statistics, Optimization & Information Computing*, 2021, **9**(3), 529-554.

[65] Ditzhaus M, Smaga Ł. Permutation test for the multivariate coefficient of variation in factorial designs. *Journal of Multivariate Analysis*, 2022, **187**, 104848.

[66] Billingsley P. *Convergence of Probability Measures*. John Wiley & Sons, 1999.

[67] Van der Vaart, A. W. *Asymptotic Statistics*, Cambridge University Press, 2000.

[68] Lo SM, Wilke RA. A copula model for dependent competing risks. *Journal of the Royal Statistical Society: Series C (Applied Statistics)*, 2010, **59**(2), 359-376.